%% file: article.tex
\documentclass[12pt]{spieman}

\usepackage[]{graphicx}
\usepackage{siunitx}
\usepackage{amsmath}
\usepackage{amssymb}
\usepackage{upgreek}
\include{units}
\usepackage[final]{changes}

\definechangesauthor[name={Julian}, color=orange]{Julian}
\setremarkmarkup{(#2)}

\title{Rubidium-traced white-light etalon calibrator for radial velocity measurements at the cm\,s$^{-1}$ level} 

\author[a]{Julian St\"urmer}
\author[a]{Andreas Seifahrt}
\author[b,c]{Christian Schwab}
\author[a]{Jacob L. Bean}

\affil[a]{Department of Astronomy \& Astrophysics, The University of Chicago, Chicago, IL 60637}
\affil[b]{Department of Physics and Astronomy, Macquarie University, Sydney, NSW 2109, Australia}
\affil[c]{Australian Astronomical Observatory, Sydney, Australia}

\begin{document}
\maketitle
\begin{abstract}
We report on the construction and testing of a vacuum-gap Fabry-P\'erot etalon calibrator for high precision radial velocity spectrographs. Our etalon is traced against a rubidium frequency standard to provide a cost effective, yet ultra-precise wavelength reference. We describe here a turn-key system working at \SIrange[]{500}{900}{\nano\meter}, ready to be installed at any current and next generation radial velocity spectrograph that requires calibration over a wide spectral bandpass. Where appropriate, we have used off-the-shelf, commercial components with demonstrated long-term performance to accelerate the development timescale of this instrument.  Our system combines for the first time the advantages of passively stabilized etalons for optical and near-infrared wavelengths with the laser-locking technique demonstrated for single-mode fiber etalons. We realize uncertainties in the position of one etalon line at the \SI{10}{\cmps} level in individual measurements taken at \SI{4}{\hertz}. When binning the data over \SI{10}{\second}, we are able to trace the etalon line with a precision of better than \SI{3}{\cmps}. We present data obtained during a week of continuous operation where we detect (and correct for) the predicted, but previously unobserved shrinking of the etalon Zerodur spacer corresponding to a shift of \SI{13}{\cmps} per day.
\end{abstract}



\section{Introduction}
\label{sec:intro}
The radial velocity method has been one of the most important observational techniques in the history of the field of exoplanet science, and it will continue to be critical for making many of the most significant exoplanet discoveries anticipated over the next two decades. One reason for this is that radial velocity surveys can efficiently reveal the census, masses, and orbits of objects in the planetary systems around nearby stars \cite{howard10, mayor11}. These data are valuable constraints on theories of planet formation and evolution. Furthermore, knowing the orbits of the exoplanets around the nearest stars will also be useful for guiding future direct imaging efforts to obtain their spectra \cite{brown15}.

Another reason for the ongoing importance of the radial velocity method is that radial velocity measurements are a way to confirm and measure the masses of planet candidates identified by transit searches and to search for additional non-transiting planets in the same systems \cite{gettel16}. Transiting planets are the only planets for which we can determine masses and radii, and examine their atmospheres. Knowing a planet's mass and radius together give us constraints on its bulk composition, and spectroscopic studies can reveal further details about the composition and physical conditions of its atmosphere. Together, bulk composition and atmospheric properties are a powerful diagnostic of a planet's origins and evolution.

One of the main goals of the field of exoplanet science is to ascertain the frequency, and the physical and orbital characteristics of low-mass planets. The identification and characterization of potentially habitable worlds is an aspect of this quest that is especially exciting. The techniques for Doppler spectroscopy have currently progressed to the point that precisions of 1 -- 2\,m\,s$^{-1}$ are routinely obtained on bright stars \cite{howard11, lovis11}, and precisions of 60 -- 80\,cm\,s$^{-1}$ have been obtained in a few select cases \cite{pepe11}. This level of precision enables the detection of Earth-mass planets with periods up to a few tens of days around solar-type stars. However, Earth imparts a radial velocity signal with a semi-amplitude of only 9\,cm\,s$^{-1}$ on the Sun when viewed edge on. Therefore, the detection of terrestrial planets in the habitable zones of Sun-like stars is still out of reach with current instruments, and the precision achievable with the radial velocity technique must be substantially improved to pursue this compelling goal.

One of the key instrumental aspects that must be improved in radial velocity measurements to reach the level of precision required for the detection of Earth analogues is the spectrograph wavelength calibration \cite{fischer16}. Following the stunning success of the HARPS instrument \cite{mayor03}, the next generation of radial velocity instruments are being designed to feature high thermo-mechanical stability. However, even at the highest levels of instrumental stability, long-term radial velocity drifts exceeding the signal for an Earth-like planet are to be expected. Therefore, the next generation of radial velocity spectrographs require a wavelength calibrator that provides both the local information density and the long-term stability to measure and track instrumental drifts with a precision of \SI{10}{\cmps} or better over time scales of minutes to years. In addition, detailed instrument characterization, e.g. to determine the stitching error of the \replaced{detector}{CCD} \cite{wilken10}, is required for sub-\si{\mps} RV measurements. All these requirements lead to the necessity for the calibration source to provide a dense grid of evenly distributed lines of uniform intensity in order to maximize the extraction of the Doppler information content over a broad wavelength range.

As of today the technology most closely fulfilling the requirements for next generation wavelength calibration is the laser frequency comb (LFC) generated by femtosecond pulsed lasers \cite{murphy07,li08,quinlan10, ycas12, phillips12}. The self-referencing of the laser-comb lines produces exceptional stability and frequency precision (better than $10^{-14}$), locked to an atomic standard for very high accuracy. A drawback of LFCs is that the comb lines have a frequency spacing of only a few 100\,MHz. Astronomical spectrographs used for radial velocity measurements usually have a resolution of 10--30\,GHz, which is insufficient to resolve the narrowly spaced comb structure. To increase the comb line spacing, a common solution is to filter out most of the lines with external Fabry-P\'erot (FP) cavities \cite{steinmetz09,quinlan10}, creating a so-called ``astro-comb''. 

Over the last decade, LFCs for astronomical applications have seen a tremendous development, culminating in the recent introduction of a commercial product offered by Menlo Systems (Germany), covering 450--1400\,nm at 25\,GHz line spacing and $<$3dB intensity variations over the entire bandpass. Precursors of this system have been developed for HARPS-S and ESPRESSO \cite{locurto12m}. However, the cost for this system is often prohibitive, and despite its advantages recent RV spectrograph projects, e.g. CARMENES/CAHA3.5m \cite{carmenes}, SPIROU/CFHT \cite{spirou}, or \replaced{KPF}{Shrek}/Keck \added[]{\cite{gibson2016}}, either do not use an LFC or include it only as a future upgrade, owing to the cost impact on the project. Therefore, a cheaper alternative for wavelength calibration of next-generation RV instruments is desirable.

A simple and elegant alternative approach to generating a comb reference spectrum is to use the cavity resonance lines of a Fabry-P\'erot (FP) etalon illuminated with white light \cite{wildi09,wildi10,wildi11,wildi12,halverson12,halverson14}. An ``ideal'' FP etalon illuminated with a broadband light source produces emission lines that are equidistant in frequency space. The position, line width, spacing, and amplitude of these synthetic lines can be easily tailored to match the spectrograph requirements by tuning the finesse and free spectral range of the etalon, which are functions of the separation and reflectivity of the cavity mirrors. 

The position of the peaks in an etalon spectrum is only a function of cavity length $l$ (essentially mirror spacing $d$, illumination angle $\Theta$, and refractive index $n$) and interference order $m$. In contrast to the LFC, the absolute frequency of the peaks of the etalon comb are not known with great accuracy, owing to the degeneracy between $l$ and $m$. However, what is important is the stability of the line positions, and this can be achieved by keeping the effective cavity length of the etalon constant. It is the lack of the extreme accuracy with which the frequency of each and every comb line is known, that makes the FP etalon inferior, but at the same time so much cheaper than a LFC. Fortunately, in the context of radial velocity measurements, seeking relative RV measurements with extreme precision, extreme stability is required, not extreme global accuracy, and the FP etalon can fulfill this requirement. Practically, however, the stability of a passively stabilized etalon can not be achieved over the desired time scales of exoplanet signals of weeks to years. Measuring and tracking of the etalon line positions with high precision becomes mandatory and we demonstrate the end-to-end implementation and performance of this method in this paper.

The FP etalon-based wavelength calibration system described here was developed for the MAROON-X instrument, which is a new fiber-fed, red-optical, high-precision radial-velocity spectrograph for \replaced{the 8.1\,m Gemini North telescope on Mauna Kea in Hawai'i}{one of the twin 6.5m Magellan Telescopes in Chile}, currently under construction at the University of Chicago \cite{maroonx}. While being developed for a particular instrument, it is suitable for calibrating any radial velocity spectrograph at the few cm\,s$^{-1}$ level after simple modifications to match the desired wavelength range and spectral resolution. Parts of the results have already been published in a SPIE proceeding \cite{stuermer2016}. In \S2 we describe the design of the FP etalon. In \S3 we describe our frequency stability monitoring of an etalon line using a scanning laser and rubidium gas cell. We present the performance of the system in \S4. We conclude with a discussion of applications and future development in \S5.
\begin{figure*}
\begin{center}
\includegraphics[width=0.52\textwidth]{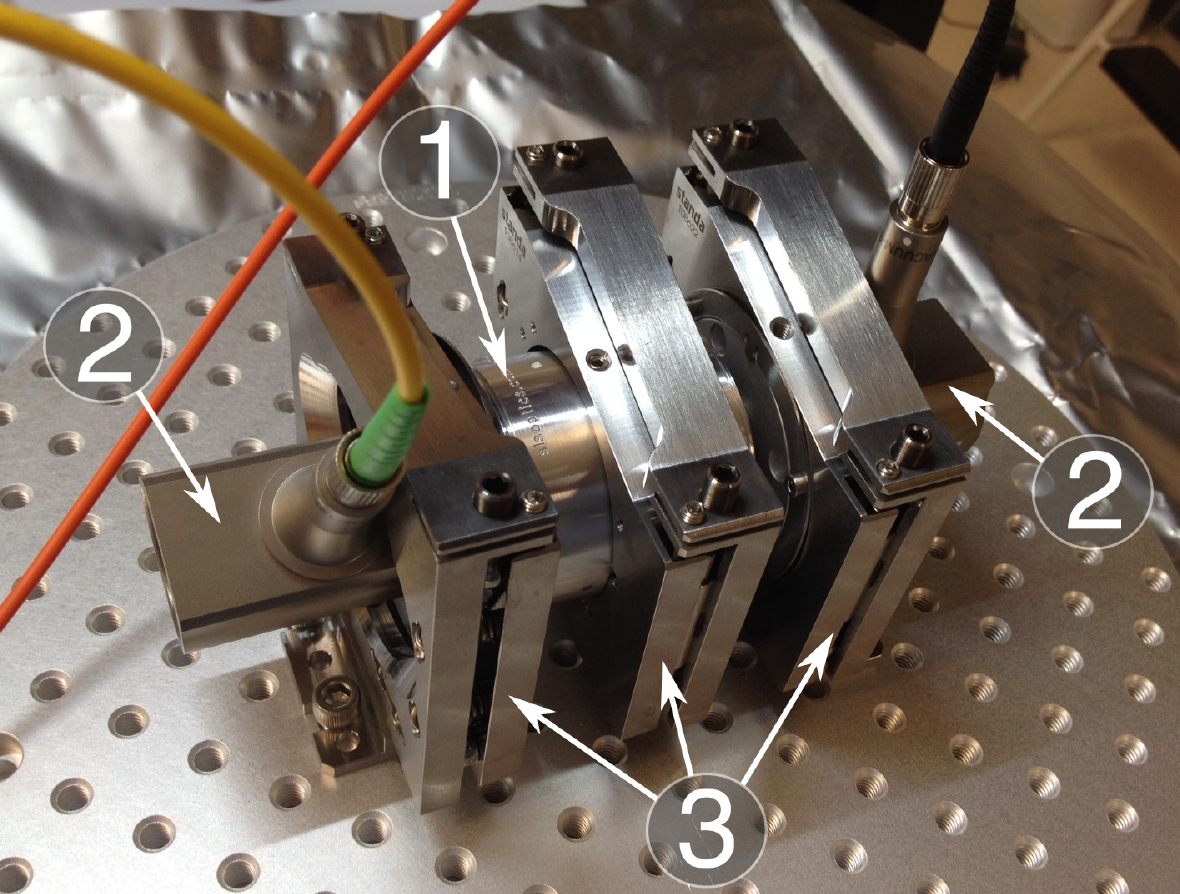}
\includegraphics[width=0.436\textwidth]{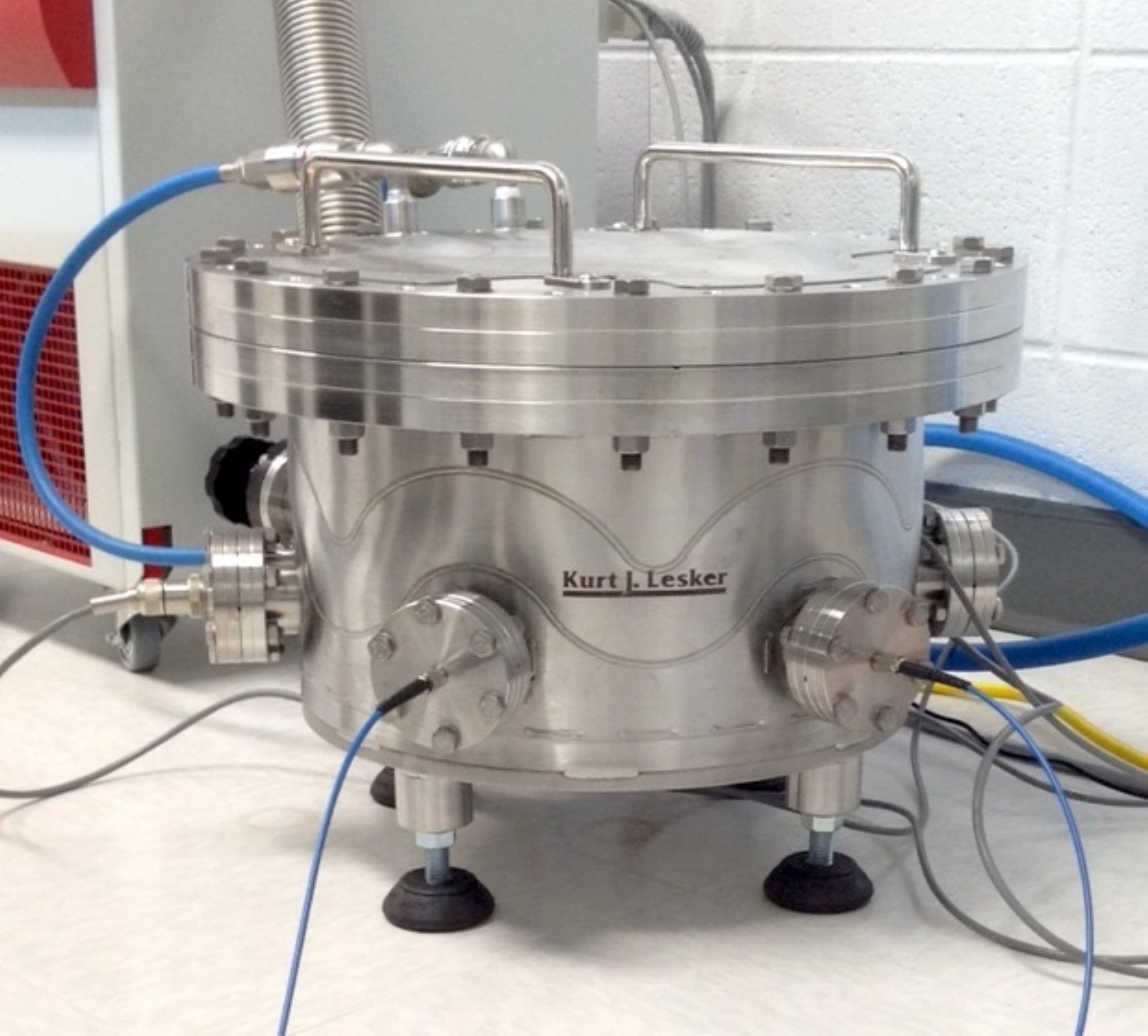}
\vspace{1mm}
\caption{{FP etalon opto-mechanics and vacuum chamber}. Left: FP etalon (1) and OAP collimators (2) in vacuum compatible tip-tilt mounts (3) on a breadboard before vacuum integration. Right: System integrated in a vacuum chamber from J.K. Lesker with in-built channels for liquid circulation to provide temperature control at the \SI{\leq5}{\milli\K} level with an external bath thermostat (not shown). During operation the vacuum vessel is contained in another insulation box (also not shown here) to attenuate temperature variation of the room.}
\label{lesker}
\end{center}
\end{figure*}

\section{Etalon Design}
The design specs for the MAROON-X FP etalon calibrator are as follows:
\begin{itemize}
\item Dense comb of lines over 500--900\,nm 
\item 15 GHz comb line spacing (on average every third resolution element for a R=80,000 spectrograph) 
\item Comb lines completely unresolved at R=80,000 
\item Intensity variations across the bandpass smaller than a factor of two
\item Line position tracked with a precision of better than \SI{10}{\cmps} over timescales of minutes to months
\item Filling the full well of the CCD in a R=80,000 spectrograph with a sampling of 3.5 pixel per resolution element and 30\% total throughput (from telescope fiber feed to spectrograph CCD) in less then one minute integration time
\item Scalable flux rate to allow exposure times up to 30\,min
\item Fiber-fed for easy integration
\item Autonomous operation over at least six months without intervention requirements 
\end{itemize}


We have purchased a 30\,mm diameter (25\,mm clear aperture) vacuum-gap FP etalon from SLS Optics (UK), with a free spectral range of 15\,GHz, and a finesse of $\sim$40. The line width of the etalon is thus about 400\,MHz (\SI{\sim300}{\mps}), completely unresolved by MAROON-X. 

We have also considered the use of a fiber FP etalon \cite{halverson12,gurevich14,schwab15}, but find that despite their compactness, they exhibit two important shortcomings. Since the interference takes place in the fiber (fused silica), they are about 300 times more sensitive to temperature changes than vacuum-gap etalons with Zerodur spacers ( \SI{0.053}{\per\milli\kelvin} vs. \SI{15}{\per\milli\kelvin} for a \SI{10}{\cmps} change, respectively \cite{schwab15}). Most importantly, the fused silica used in fibers has a notable birefringence, making the effective position of the etalon lines polarization dependent. A strict control and management of the polarization of the injected light is necessary to reach a sub-\si{\mps} level of stability \cite{halverson12,gurevich14}. We thus concluded that a vacuum-gap FP etalon is currently the least risky choice.

The spacer in our etalon is made of \SI{10.01}{\milli\metre} Zerodur Grade 0. The FP etalon coatings have a reflectively of $R\sim94\%$ over 500--900\,nm. broadband anti-reflection (BBAR) coatings on the outside of the etalon have a reflectivity of $R\leq0.5\%$ over this wavelength range. The etalon is mounted stress-free in a vacuum compatible aluminum cylinder, as it needs to be kept in a vacuum chamber to facilitate pressure and temperature stability to better than \SI{e-3}{mbar} and \SI{10}{\milli\kelvin}, respectively.

The etalon is fed by a customized (vacuum compatible) off-axis paraboloid (OAP) fiber collimator from Thorlabs with \SI{2}{"} focal length. An identical OAP collects the light from the etalon and focuses it in on the output fiber. Both OAP collimators as well as the etalon are mounted in vacuum compatible stainless-steel tip-tilt mounts from Standa (Lithuania) on a circular vacuum-compatible aluminum breadboard (see Figure\,\ref{lesker}).

After extensive testing we decided to feed the etalon with a single-mode (SM) fiber (Thorlabs SM600) for two reasons. First, despite our best efforts to stabilize the illumination of the etalon we could never retrieve a clean etalon line profile with our laser scanning technique (see below) when using a multi-mode (MM) fiber feed. Small deviations in the angular distribution of the light, introduced by modal noise in the fiber, insufficient opto-mechanical stabilization of the fiber feed, and changing focal-ratio-degradation (FRD) effects led to asymmetries and shifts of the etalon lines since the effective cavity length changes with the angle of the collimated beam. Models suggest that a beam deflection of only \SI{5}{\arcsecond} in the etalon will shift the barycenter of the etalon lines by \SI{10}{\cmps}. This translates into extreme sensitivity of the etalon to non-uniform illumination for MM fibers. 

Second, our laser scanning technique (see below) relies on the fact that the laser light and the white light share the exact same optical path and mode volume in the etalon. This is impossible to achieve with MM fibers. By using a SM fiber for illumination of the etalon, we could eliminate both problems at the cost of reducing the throughput by a factor of about 300 compared to a \SI{50}{\micron} multi-mode fiber illuminated at its full NA at \SI{650}{\nano\metre}. We have measured the spectral throughput behavior of the SM fiber over the \SIrange[]{500}{900}{\nano\metre} bandpass and found that slightly coiling a short piece of the SM fiber suppresses unwanted higher modes at the blue end of our bandpass while keeping the mode cut-off just \replaced{beyond}{red of} \SI{900}{\nano\metre}.

Thanks to the excellent alignment of the etalon achievable with the laser scanning technique and its high peak transmission (\SIrange{60}{80}{\%}), we can still retrieve enough photons from a laser-driven white light source (Energetiq LDLS model EQ-99XFC) to fulfill our comb brightness requirement. Based on the measured efficiency of the LDLS, the measured throughput of our etalon, and the coupling losses of the fibers and collimators, we conservatively estimate that we will still receive an average flux of \SI{\sim1.5e7}{photons\per\second} per etalon line. Assuming only 30\% end-to-end spectrograph efficiency and a sampling of \SI{3.5x10}{pixel} per resolution element, we estimate that the detector will still record over 600\,ke- per minute per pixel for each etalon line from a \SI{50x150}{\micron} calibration fiber.

We use a dichroic (Thorlabs DMLP950) to select a bandpass of \SIrange[]{400}{900}{nm} from the LDLS in order to suppress the UV and NIR part of the LDLS which can lead to unwanted thermal effects in the etalon. At the output of the etalon, we use a standard \SI{50}{\micron} MM fiber to retrieve the comb light. All our fiber connections have at least one FC/APC connector to reduce the high-frequency fringing which otherwise occurs in short fiber runs. A color filter will be used to balance the slope in the throughput of the SM fiber and to match the brightness distribution of the etalon comb to the efficiency of the spectrograph.

The etalon assembly is mounted inside a stainless steel vacuum chamber with in-built liquid circulation channels by J.K. Lesker. Six 2.75" CF ports are used for pumping, electrical feedthrough, and custom fiber feedthroughs built in-house. After initial pumping, we achieve vacuum levels of \SI{<2e-6}{\milli\bar} with a \SI{0.5}{\litre\per\second} ion pump.
A compact cooling bath circulation thermostat (PolyScience PD07R-40) provides a temperature stability of \SI{\leq2}{\milli\kelvin} of the cooling liquid (DI water). Two PT100 sensors and a Lakeshore controller monitor the breadboard and etalon temperature inside the tank. We keep the vacuum tank inside a passive foam insulation box to further improve the temperature stability.

\section{Frequency Stability Control}

\subsection{Background}

The challenge for a passively stabilized FP etalon is that frequency drifts can occur for many reasons, including aging of materials, like Zerodur spacers in vacuum-gap etalons \cite{wildi09}, long-term degradation of the dielectric mirror coatings, changing environmental conditions (temperature and pressure), as well as changes in the illumination (i.e., thermo-mechanical stability of the light injection). Monitoring and, if desired, active control of a FP etalon tailored for astronomical research at sub-m\,s$^{-1}$ accuracy thus requires precise metrology.

\begin{figure*}[t!]
\centering
\includegraphics[trim={0 0 0 1.5cm},width=0.99\textwidth,clip]{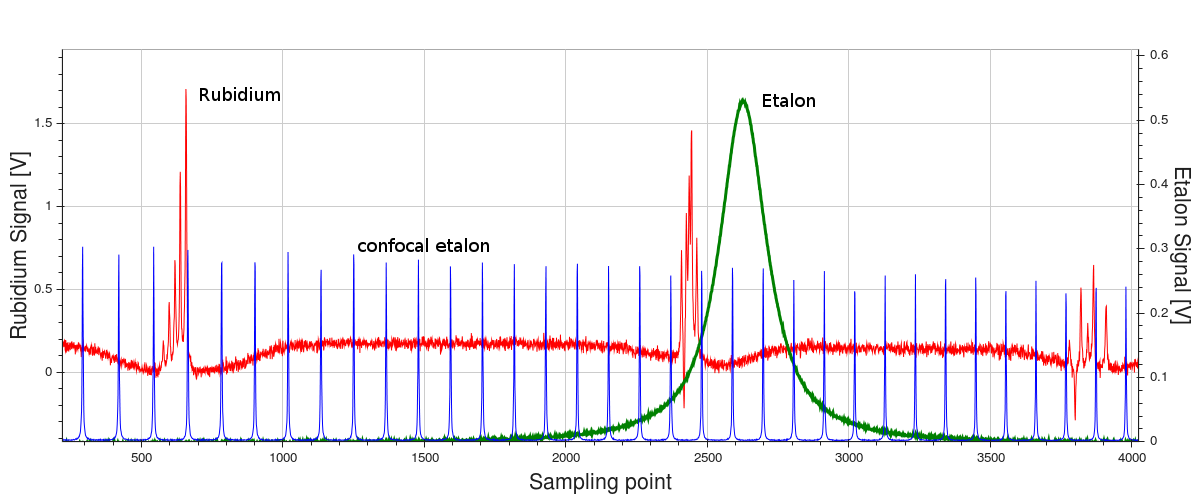}
\vspace{1mm}
\caption{{Snapshot of a single ECDL scan of our FP etalon line} (green), the rubidium reference spectrum (red), and the confocal etalon used to linearize the scan axis (blue). The scan spans approximately \SI{6.6}{\GHz} or \SI{5}{\kmps}. The rubidium transitions are hyper-fine lines of Rb$^{85}$ and Rb$^{87}$ D2.}
\label{scan}
\end{figure*}

A simple solution is to cross-check the calibration spectrum of the FP etalon against classical calibration methods, like hollow-cathode lamps (HCLs), using the RV spectrograph during daytime \cite{wildi12}. Unfortunately, this strategy is limited by the accuracy of the HCLs and poses the proverbial chicken-and-egg problem, i.e., one needs the spectrograph to calibrate the calibrator and it may be unclear whether a change in the etalon spectrum is due to the etalon itself or to changes in the HCL and the spectrograph. The ultimate limitation of this approach is the practically exploitable radial velocity content of the HCL, which in the case of ThAr is at the \SI{20}{\cmps} level \cite{pepe11}. Consequently, stabilities of \SI{10}{\cmps} over one night and \SI{1}{\mps} over 60 days were reported for a {passively} stabilized FP etalon when being measured against a ThAr lamp on the HARPS instrument \cite{wildi11}. This level of precision is not enough for the detection of Earth-like planets, where final instrumental uncertainties must be well below \SI{1}{\mps} for M-dwarfs and below \SI{10}{\cmps} for Sun-like stars, respectively. The stability of the calibrator must be several times as good to reach this limit.

The solution to this problem is to track the position of the FP etalon lines by referencing them against an absolute scale, for example, to an atomic transition. This solution eliminates the need for comparing the FP etalon signal to emission lines on the CCD of the spectrograph and provides absolute calibration of \replaced{a single FP transmission line}{the FP etalon zero point} during observations at a level of precision unmatched by other techniques, exceeded only by the laser frequency comb itself. Such a system shares many advantages with the optical frequency comb but can be built entirely with simpler and less expensive components.

Various implementations of this idea have been discussed in the recent literature \cite{gurevich14, mccracken14, reiners14, schwab15}. The proposed realizations typically use a frequency stabilized diode laser in conjunction with a rubidium gas cell, the latter serving as an absolute \deleted[]{, NIST-traceable,} frequency reference providing a relative frequency accuracy of 10$^{-11}$ \cite{steck10,steck10b}. The long-term stability of the FP etalon is then directly tied to the long-term stability of the atomic reference lines. Also, the rubidium lines can shift due to temperature (\SI{\sim4}{\kilo\hertz\per\kelvin}), laser power fluctuation (\SI{\sim13}{\kilo\hertz\per\mu\watt}) and magnetic field  (\SI{\sim1.2}{\kilo\hertz\per\mu\tesla})\cite{affolderbach}. These are not limiting the achievable stability since the temperature, power, and magnetic field can be stabilized to better than \SI{0.1}{\kelvin}, \SI{5}{\mu\watt}, and \SI{1}{\mu\tesla} over all relevant timescales. Expected systematic drifts of the rubidium reference are thus well below \SI{10}{\cmps} \added[]{, dominated by the contribution of laser power fluctuations.}

Two general approaches can be taken to tie the FP etalon to the rubidium reference. Locking techniques, common practice in laser physics, use a piezo in the FP etalon to modulate the cavity length and lock one line of the FP etalon to the laser frequency which in turn is locked to the rubidium standard \cite{mccracken14}. Alternatively, two tunable lasers, one locked to a rubidium transition, the other to the FP etalon produce a RF beat frequency that is used to measure the drift of the etalon \cite{reiners14}. The locking precision is a function of the line width and the relatively broad lines of a low-finesse FP etalon limit the achievable locking precision to \SI{\sim30}{\cmps}. The use of dual-finesse FP etalons have thus been suggested, with the low-finesse part ($\mathcal{F}\sim10$) for comb generation, and the high-finesse part ($\mathcal{F}\sim200$) for frequency locking \cite{reiners14}.

A much easier approach is to use a single tunable laser to scan simultaneously over the D$_2$ hyper-fine transition of rubidium in a saturated absorption spectroscopy setup and over a full line of a FP etalon to measure a single etalon line at extremely high resolution and accuracy against the rubidium spectrum. Using a fiber FP etalon, a radial velocity stability at the \SI{3}{\cmps} level (10$^{-10}$ relative precision) over timescales of hours was reported with this setup \cite{gurevich14,schwab15}. The error signal from the measured etalon drift against the rubidium reference can either be used to actively stabilize the etalon (e.g., via minute temperature corrections \cite{gurevich14}) or to track the zero point of a passively stabilized etalon at the \si{\cmps} level, allowing for a correction of the observed calibration comb spectrum during data analysis. This scanning technique is only limited by the ability to correct for non-linearities in the scan axis of the tunable laser and by the achievable scan speed. The latter limits the frequency with which the etalon position can be determined and can potentially limit the intrinsic precision if significant disturbances occur at frequencies higher than the scanning rate.

\subsection{Laser Scanning Setup}

Our setup follows the aforementioned approach by simultaneously scanning one etalon line and several groups of hyperfine transitions of rubidium with a tunable laser at \SI{780.24}{\nano\metre}. We use an off-the-shelf grating stabilized external cavity diode laser (ECDL) and digital controller from Toptica (DLC-100 Pro), offering a \SI{>30}{\GHz} mode-hop free scan range, in conjunction with a CoSy rubidium cell and controller unit from TEM Messtechnik (Germany). Our whole setup is fiber coupled. We outfit the rubidium cell with an additional digital temperature sensor to monitor the independently controlled temperature of the rubidium cell.  

In order to suppress stray magnetic field from surrounding equipment, most notably the strong magnetic field from the double optical isolator in the Toptica laser unit itself, we placed the CoSy unit in an off-the-shelf three layer MuMetal ZG-3 Zero-Gauss chamber that suppresses magnetic fields by a factor of $\sim1500$. While static magnetic fields are unproblematic, temporal changes in the magnetic field cause changes in the rubidium spectrum due to (unresolved) Zeeman splitting and thus potential long-term drifts of our reference frequencies. The change in the Earth's magnetic field is location dependent but ranges only from \SI{\sim65}{\nano\tesla\per year} (La Serena, Chile) to \SI{\sim116}{\nano\tesla\per year} (Chicago) and, hence, does not pose a problem even over decades. Changes in strong stray fields can, however, be a significant limitation to the long-term stability of the rubidium reference. We have thus installed a magnetic field sensor (HMC5883L) next to the rubidium cell inside the Zero-Gauss chamber to monitor residual magnetic fields with a \SI{0.2}{\mu\tesla} resolution.

We scan the laser at a scan rate of \SI{4}{\hertz} over a spectral range of \SI{6.6}{\giga\hertz} (\SI{\approx5}{\kmps}) to simultaneously measure one etalon line and three transition groups of Rb$^{85}$ and Rb$^{87}$ D2. As our scan range is relatively large\replaced{,}{and} the piezo tuning of the ECDL grating suffers from non-linearities and hysteresis effects. To compensate for this problem, we use an additional confocal etalon, similar to the one used by others\cite{gurevich14}, but with a smaller FSR (\SI{188}{\MHz}), that provides a dense comb of about 30 lines over our scan range. The confocal etalon is essential in allowing us to linearize the scan axis and to tie the localized frequency information of the rubidium reference lines to the much broader etalon line (see Figure\,\ref{scan}).

We use a small peak power of the ECDL to retrieve narrow rubidium line-widths. This allows us to resolve the previously blended lines of the  Rb$^{85}$ F2 group near the peak position of the FP etalon. Instead of the in-built scan function of the Toptica controller we use a LabJack T7 DAQ to generate a smoothed saw-tooth ramp to drive the piezo scan. We use the same DAQ to digitize the output signal of several photodiodes to produce the four scan signals (FP etalon, confocal etalon, rubidium reference, and laser power) with 4096 sampling points in real-time. 

\begin{figure*}
\centering
\includegraphics[bb=0 0 612 792, width=6in,clip]{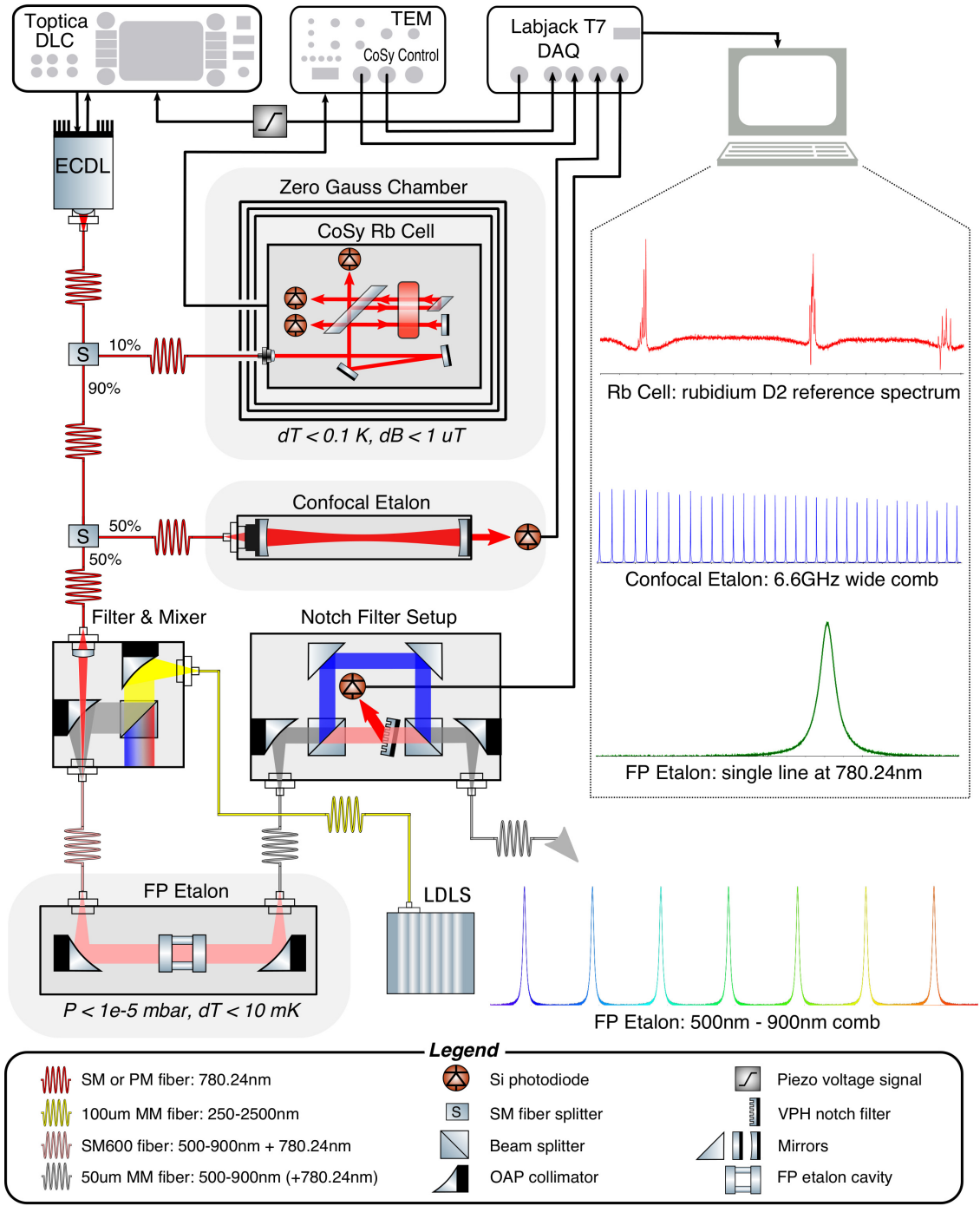}
\caption{{Schematic of our calibrator setup}. The external-cavity diode laser (ECDL) scans simultaneously a rubidium gas call, a confocal etalon, and our FP etalon. Laser light and filtered white light from a laser-driven white light source (LDLS) are combined (see ``Filter \& Mixer'') into a SM fiber to illuminate the FP etalon. The laser light scanning a single FP etalon line is removed from the comb with a VPH notch filter and analyzed with a photo-diode (see ``Notch Filter Setup''). The result of the process is a comb of lines spanning 500--900\,nm with a FSR of \SI{15}{\GHz} and a line width of under \SI{400}{\MHz} anchored to the rubidium reference with a precision of better than \SI{3}{\cmps}.}\label{schematic}
\end{figure*}

We combine the filtered white light from the \replaced{LDLS}{LDSL} and the laser light into the SM600 input fiber with two OAPs and a fiber collimator. At the end of the output fiber of the etalon (before coupling into the spectrograph) we send the light through an Ondax SureBlock ultra narrow-band notch filter to remove the laser line from the white light comb and reflect it onto a photo diode for measurement. The OD of the notch filter is about 4 and the FWHM is \SI{0.45}{\nano\metre}. This is sufficient to block the laser light and only lose etalon comb lines for about \SI{5}{\%} of the FSR in order 79 of the spectrograph. The beam has to be re-collimated for the notch filter and we use a setup with two off-the-shelf OAP collimators from Thorlabs with an EFL of \SI{7}{\milli\metre} to keep the collimated beam size small enough for the notch filter and the photo diode. A schematic of the complete setup is shown in Figure\,\ref{schematic}.

\subsection{Laser Scanning Software}

We have developed a Python software package that provides fast and robust fitting routines and a sophisticated monitoring system. By using a modular design the software is easy to maintain. The work flow of the software is as follows:
\begin{enumerate}
\item The LabJack DAQ synchronously writes the piezo signal for the laser scan, reads the rubidium, laser power, and both etalon signals in real time and sends a data stream back to the host computer.
\item Our software fits a Moffat function to each confocal etalon peak and a sum of six Lorentzian peaks plus a Gaussian background for each of the three rubidium groups. 
\item Using the measured confocal etalon peak positions together with the measured rubidium positions and their known absolute frequencies, we linearize and calibrate our scan axis in units of absolute frequency. We skip the first few hundred sampling points of the scan to remove the down-slope of the saw-tooth scan pattern and the strongest non-linearities. We then use a univariate spline with a fixed number of nodes to describe the remaining non-linearities of the scan axis induced by the hysteresis of the piezo in the ECDL. 
\item Using the linearized scan axis in units of absolute frequency, we fit a Lorentzian to the etalon peak and retrieve the absolute position of our etalon peak in THz.
\end{enumerate}

\begin{figure*}[ht!]
\centering
\includegraphics[trim={0 0 0 0cm},width=0.99\textwidth,clip]{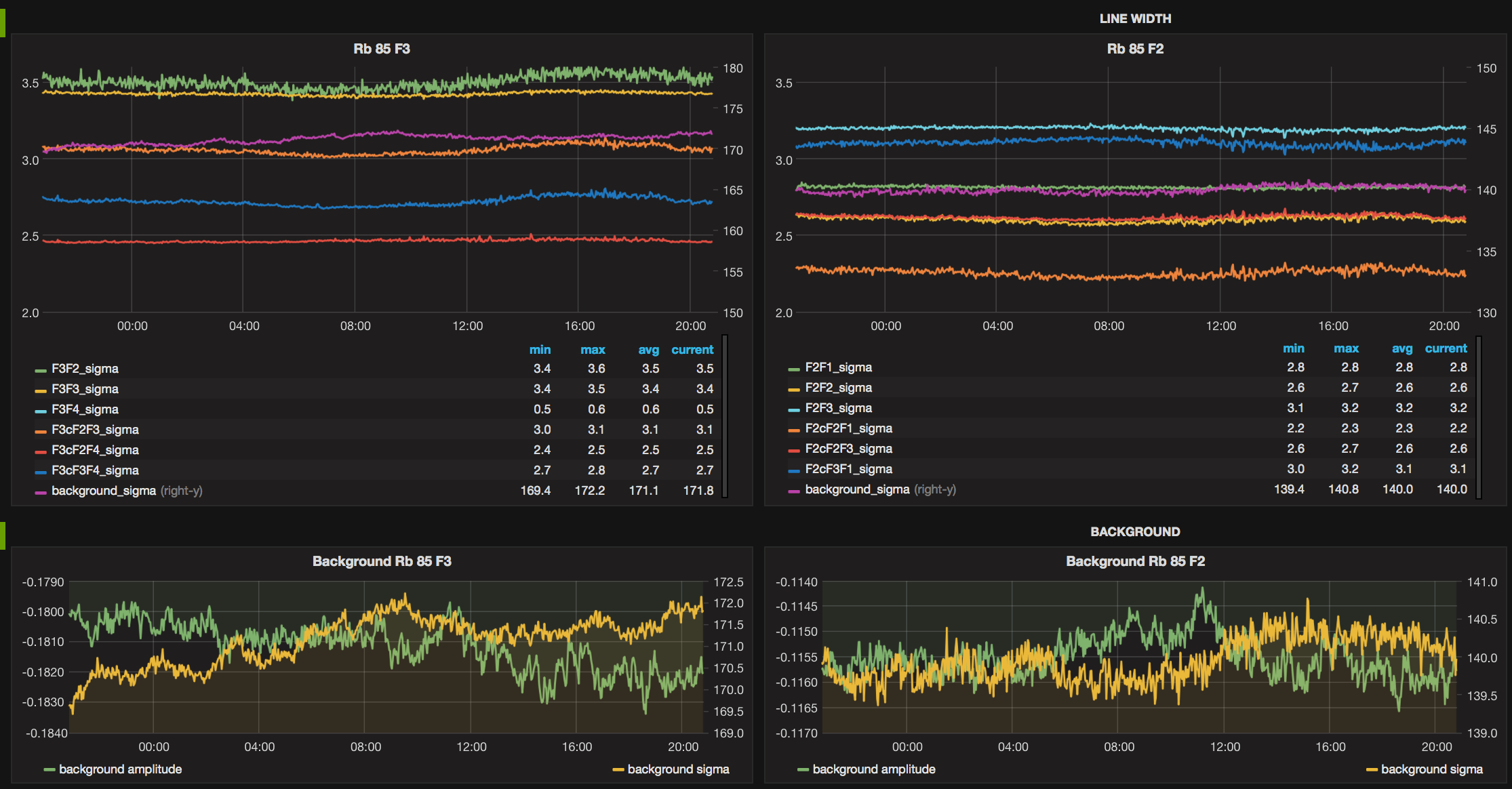}
\vspace{1mm}
\caption{{Snapshot from the Grafana web interface} showing details of the rubidium line statistics for a 24\,hr period. \replaced{Each column contains data for two of the three rubidium groups shown in Figure\,\ref{scan}, namely Rb$^{85}$\,5S$_{1/2}$\,F=3 and Rb$^{85}$\,5S$_{1/2}$\,F=2. The top row shows the line width, the center row the amplitude and width of the Gaussian (velocity broadened) background signal.}{ Each column contains data for the three rubidium groups shown in Figure\,\ref{scan}, namely Rb$^{85}$\,5S$_{1/2}$\,F=3 and, Rb$^{85}$\,5S$_{1/2}$\,F=2, and Rb$^{87}$\,5S$_{1/2}$\,F=1, respectively. The top row shows the line width, the center row the amplitude and width of the Gaussian (velocity broadened) background signal, and the bottom row the fit residuals of the rubidium line positions to the linearized frequency axis. The light blue curve in the lower left panel shows a weak rubidium line near SNR$\leq$5 that is not included in the solution of the frequency scan axis and plotted for monitoring purposes only.} A large number of such graphs can be quickly configured on dashboards, allowing for a convenient and efficient monitoring of the health and performance of the system.}
\label{grafana}
\end{figure*}

In addition to receiving and fitting the signals and steering the piezo, the software also reads out a number of environmental sensors (room pressure, room temperature, etalon vacuum level, etalon temperatures, rubidium cell temperature, magnetic field strength, etc.). 

All sensor data along with the fitting parameters are written to an InfluxDB database. This enables an efficient storage of a large number of datasets and it provides an easy interface to visualization tools like Grafana (see Figure\,\ref{grafana}). Easy to use (and live) visualization of the fit parameters turned out to be indispensable for debugging and optimizing the setup and provides a convenient way to monitor our system locally as well as remotely. 

For example, the live-scan of an etalon line during adjustment of the collimators allows us to perfectly align the beam by adjusting the tip-tilt of the etalon mount such that the etalon line is locked in its frequency minimum. This assures that the collimated beam angle $\Theta$ is zero to within one arcmin or less. Given the $\cos(\Theta)$ dependence of the etalon line positions and, hence, the decreased illumination angle stability $d\lambda/d\Theta$ for $\Theta>0$, a near-perfect alignment greatly reduces the susceptibility of the etalon to thermo-mechanical and illumination instabilities. \added[]{Note that the exact angular beam alignment also ensures that we always monitor the same interference peak for fixed temperature and pressure conditions. Although no absolute order number can be assigned to the peak, it can still be uniquely identified.} Watching an etalon line ``live'' with a \SI{4}{\Hz} update rate also allows us to identify instabilities and minute disturbances in real time.

Room pressure variations lead to a slow drift of our scan range (due to $dn/dT$ variations on the ECDL grating and  $dl/dT$ variations of the cavity length between the grating and the diode) and would eventually shift the rubidium groups out of our scan range. To avoid this, we run a PID loop on the average rubidium position and change the piezo offset voltage to keep the scan range fixed in frequency space. Individual corrections of the piezo offset voltage are kept small enough to avoid jumps and additional non-linearities during a single scan. Free fit parameters to the frequency axis in step 3 above include the zeropoint and the FSR of the confocal etalon. We are thus only relying on the constant spacing of the confocal etalon lines, not on their
absolute position. This greatly minimizes our requirements to thermally and mechanically stabilize the confocal etalon. In fact, we allow the confocal etalon to drift with temperature and pressure as these drifts are smaller than other noise sources over a single scan (of \SI{0.25}{\second}). 

Last but not least, laser power fluctuations have an important impact on the achievable precision and long-term stability if not properly corrected for. On short time scales, i.e., over a single scan, the laser power changes monotonically by 10\%--20\% as a function of ECDL grating position. This feature can not be fully removed, even with an optimized setting of the feed forward current, as it is non-linear over the scan range. This systematic power variation has two impacts: It skews the measured line profile of the etalon peak and introduces systematic frequency differences between the three rubidium groups in the scan. When the slope in the laser power is constant, this is of no concern, but if this function changes over time, it will shift the measured vs. the true etalon line position as well as the reference frequency. Since individual rubidium lines and the lines of the confocal etalon are extremely narrow, the slope (and its potential change) over an individual line is negligible and might only increase the residuals when fitting a frequency solution to the scan axis (step 3 above). Long-term drifts of the mean laser power incident on the rubidium gas will have an impact on all the reference line position and hence the mean reference frequency. To account for these limitations, we monitor the laser power during each scan to identify, and correct for, changes, both in the laser power distribution over the scan and the average power on longer timescales.

\section{System Performance}

We measure a FWHM of \SI{\sim340}{\mega\hertz} for our scanned etalon line and hence a local finesse of almost $\mathcal{F}=44$. This is slightly higher than the theoretical prediction by SLS ($\mathcal{F}=39$ at \SI{780}{\nano\metre}). We attribute this to the smaller beam diameter (\SI{\sim12}{mm} $1/e^2$ diameter) compared to the CA used by SLS to calculate the finesse (\SI{\sim25}{mm}).

\begin{figure}
\begin{minipage}[c]{0.49\textwidth}
\includegraphics[width=0.99\linewidth,clip]{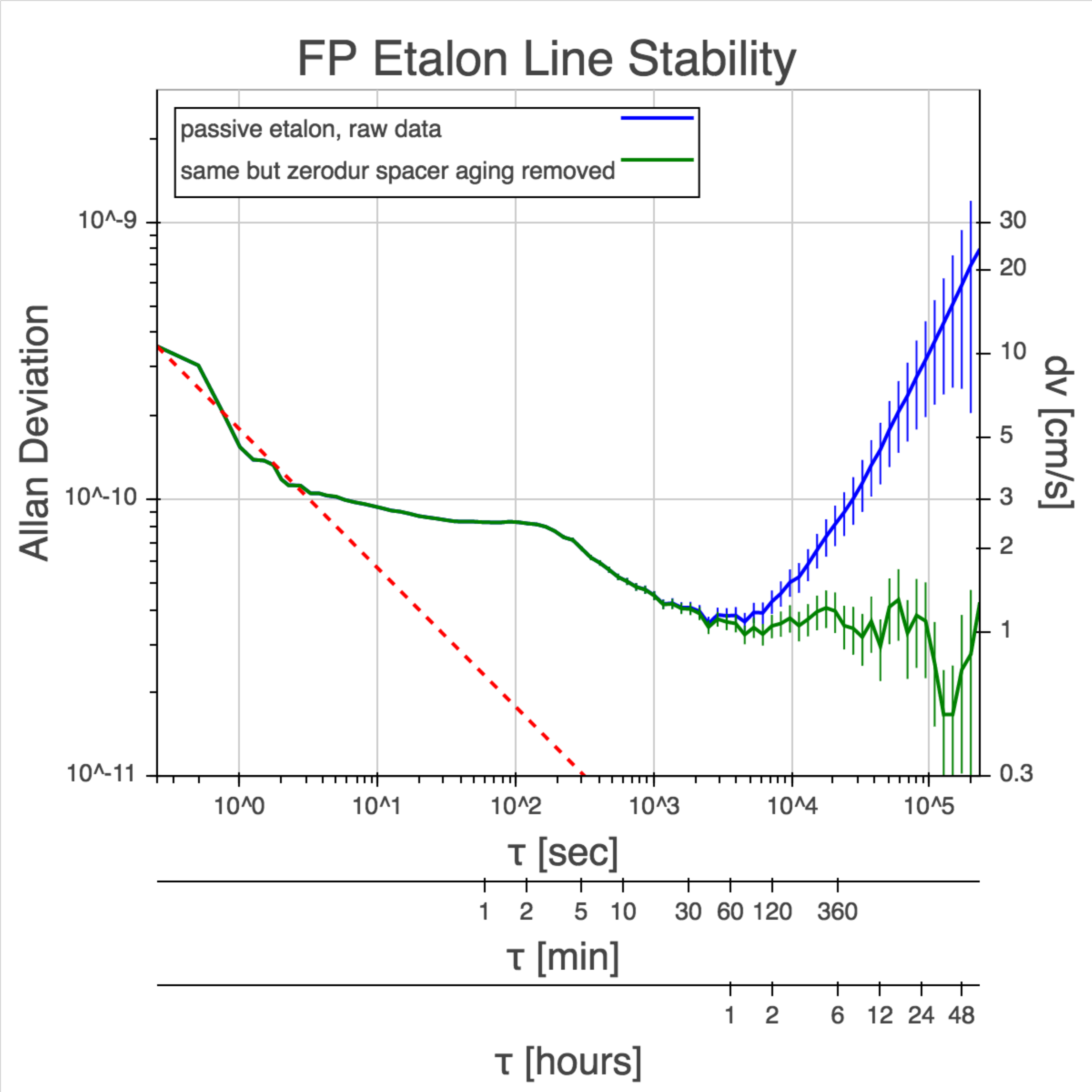}
\end{minipage}
\begin{minipage}[c]{0.49\textwidth}
\caption{Allan deviation plot for our Fabry-P\'erot etalon over one week of continuous operation. This plot shows the residuals for a single etalon line measured against the rubidium reference for a completely uncorrected, passively stabilized etalon (blue). If we subtract a linear trend of \SI{13}{\cmps}/day to account for the shrinkage (aging) of the Zerodur spacer, we reach \SI{\sim1}{\cmps} precision per etalon line when binning in 60\,min intervals (green). The red dashed line shows the expected binning behavior for white noise distributed uncertainties ($1/\sqrt{\tau}$). Error bars express the statistical uncertainty in the Allan deviation and are calculated from the number of pairs used to compute the Allen deviation at a certain binning period (adev/$\sqrt{n}$). }\label{allan}
\end{minipage}
\end{figure}

The frequency of the etalon line shows a standard deviation about \SI{\sim150}{\kilo\hertz} (\SI{\sim12}{\cmps}) \replaced{between}{among} single scans. Further binning, particularly over typical integration timescales of RV spectrographs\added[]{, ranging from a few seconds to 30 minutes,} will \replaced{reduce}{improve} the uncertainty in the etalon position. In Figure\,\ref{allan} we show the Allan deviation plot for one week of continuous operation of our etalon. After less then 10\,seconds of binning, the Allen deviation drops to under \SI{3}{\cmps} but then falls slower than expected if deviations would be Gaussian distributed and no systematic \added[]{errors} would be present (white noise case). In fact we find a local maximum around 2\,minutes of binning before the Allan deviation drops again to about \SI{\sim1.5}{\cmps} for data binned over 60\,minutes. 

For longer timescales we find that we are dominated by a linear drift of about \SI{13}{\cmps}/day, which we identify as the aging of the Zerodur spacer in the etalon, in good agreement to known shrinkage rates of Zerodur \cite{zerodur}. If we subtract this drift, we end up with Allan deviations of \SI{\sim1}{\cmps} for binning periods longer than 60\,min.

It should be noted that the Allan deviation for binning intervals longer than about \SI{30}{\min} is not a relevant metric in the context of precise RVs. When our rubidium-traced etalon comb is used as a simultaneous calibration source, the precision of the measured comb line position, averaged over the typical exposure times of the stellar spectra, is the important metric for the achievable long-term performance. For such timespans (and, hence, permissible binning intervals) of \SI{10}{\second} to \SI{30}{\min}, we can determine the position of our etalon comb to \SI{\leq3}{\cmps} relative to the rubidium reference and track the comb drifts with this precision at any given time.

\begin{figure}

\begin{minipage}[c]{0.99\textwidth}

\includegraphics[width=0.99\linewidth,clip]{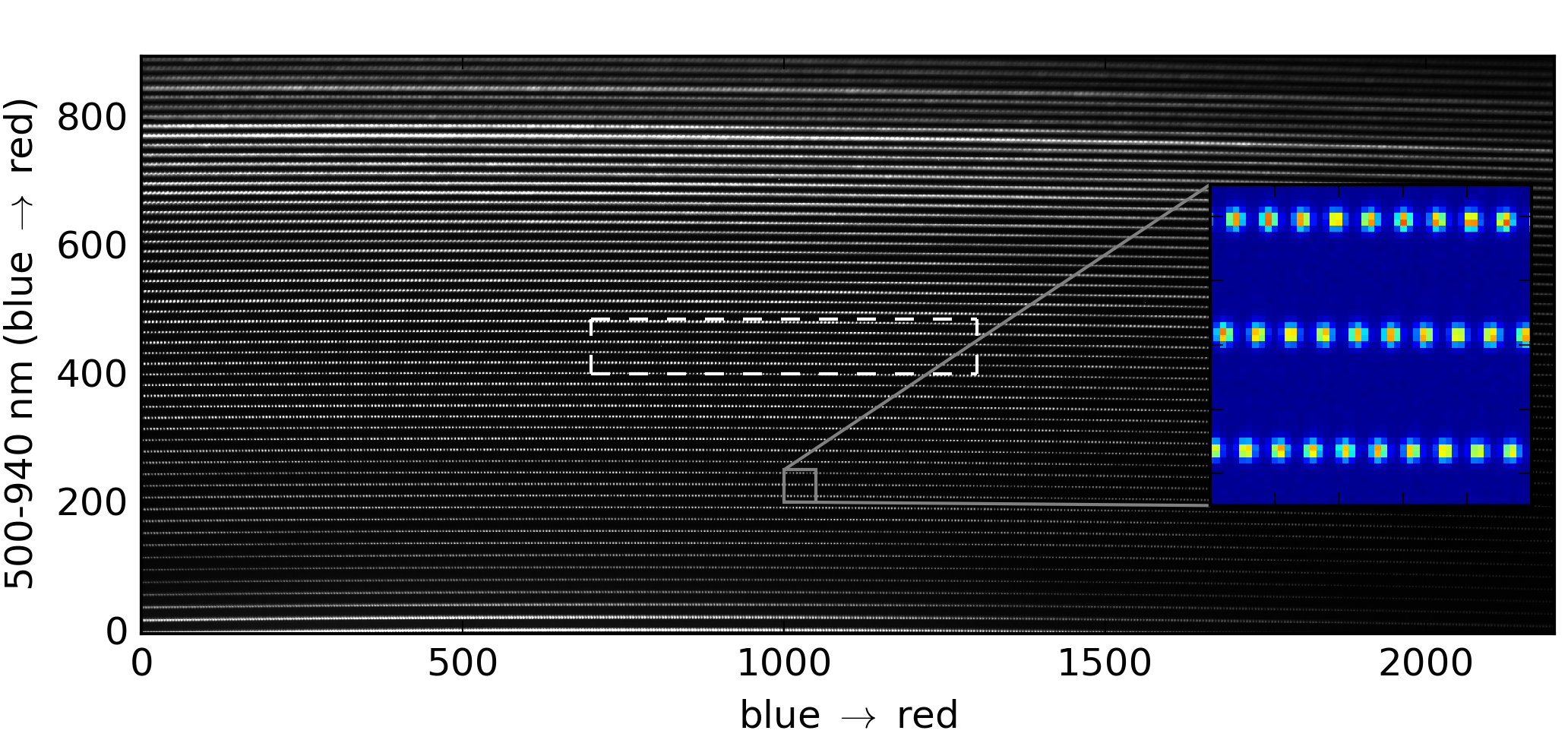}
\end{minipage}\\
\begin{minipage}[c]{0.99\textwidth}
\includegraphics[width=0.99\linewidth,clip]{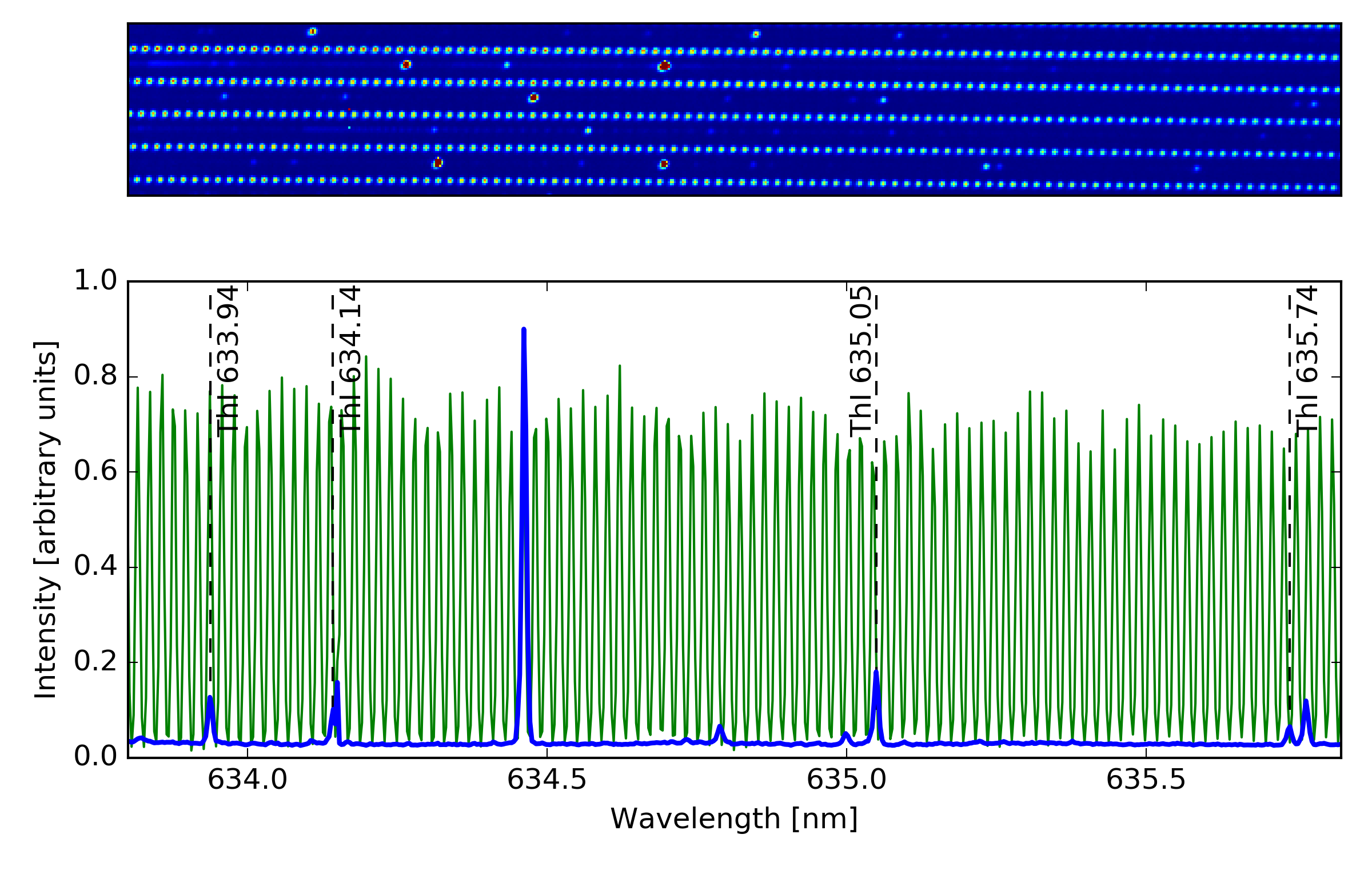}
\end{minipage}\\
\begin{minipage}[c]{0.99\textwidth}
\caption{Top: Raw etalon spectrum from \SIrange[]{500}{940}{\nano\meter} recorded with a modified RHEA spectrograph, spanning the MAROON-X wavelength range. Note that the aliasing pattern on the 2D-image is due to downsampling of the image and not present in the full frame data. Similarly, intensity variations in the image, particularly in the \SIrange[]{600}{800}{\nano\meter} range, are mainly caused by image downsampling effects coupled with the sudden decrease in image quality beyond 650 nm. Center and bottom: 2D and extracted etalon and ThAr spectrum (from the region marked by the dashed box in the full 2D image above), highlighting the increased line density and smaller peak intensity variations of the etalon comb in comparison to the ThAr spectrum.}\label{etalon2d}
\end{minipage}
\end{figure}

To verify the expected broad band performance, particularly the single-mode throughput of the SM fiber over our \SIrange[]{500}{900}{\nano\meter} bandpass \deleted[]{and the modal noise behaviour}, we recorded etalon spectra with a modified RHEA spectrograph \cite{Feger2016}. RHEA is a SM fiber fed spectrograph designed to work from \SIrange[]{450}{650}{\nano\meter} and delivers a spectral resolving power of $\sim$ 72,000 at \SI{550}{\nano\meter}. By replacing the original cross disperser with a lower dispersion prism, we extended the wavelength range to almost \SI{940}{\nano\meter} at the cost of poor image quality in the red. Still, we were able to verify the comb-like spectrum over the whole design wavelength range (see Figure \, \ref{etalon2d}). Strong modal noise was present in the spectrum, modulating the intensity of etalon comb spectrum over individual orders. \added[]{Note that the modal noise occurred, because we had to couple the light from the \SI{50}{\micron} output fiber of the etalon back into a SM fiber to achieve the full resolution of RHEA.} After employing a General Photonics dynamic multi-mode scrambler, the modal noise pattern is strongly attenuated. Additional active mode scrambling is built into the fiber feed of MAROON-X, further suppressing any residual model noise effects.

\section{Discussion}
Our FP etalon calibrator combines for the first time the advantages of passively stabilized vacuum-gap bulk etalons \cite{wildi12} with the laser-locking technique demonstrated for single-mode fiber etalons \cite{gurevich14,schwab15}. By incorporating the advantages of  illumination stability provided by a single mode fiber feed while avoiding the birefringence effects in SM etalons, we have created a calibrator that meets the needs of current and next generation RV spectrographs. The laser scanning technique allows us to use a FP etalon with moderate finesse for simultaneous comb creation and zero point tracking with high precision. Effective filtering of the laser light from the resulting etalon comb allows the usage of reference frequencies within the bandpass of the etalon comb without contamination of the white light comb or significant loss in comb coverage.

We have demonstrated that we can measure the line position of our etalon with a precision of \SI{\leq3}{\cmps} relative to the rubidium reference during every exposure of a RV spectrograph. We can track the long-term drifts of the etalon with this precision at any time, removing any high-level intrinsic stability requirements on the etalon line position for timescales longer than a few seconds. In fact we can measure the \SI{\sim13}{\cmps} per day drift from aging of the etalon Zerodur spacer and apply a correction to the comb spectrum at the data reduction level. \deleted[]{ Due to the Zerodur aging alone, a passively stabilized etalon without referencing to rubidium would become useless as a primary wavelength calibrator after less than a day, if a stability of the calibration source at the $\SI{10}{\cmps}$ level is required.}

The currently achieved level of precision is exceeding our requirements for MAROON-X, but we still see room for further improvement. As of this writing we have yet to thoroughly characterize the etalon as well as the rubidium reference in terms of its as-built behavior with respect to environmental changes compared to the theoretical expectations. This will ultimately improve the precision of the measured etalon line position but will also assure the long term absolute stability of our rubidium reference. 

For example, we find the stability of the mean laser power injected into the CoSy setup (\SI{\sim300}{\mu\watt}) to be just under \SI{1.7}{\%} P-V over the course of a week, which is already approaching our long term stability limit of \SI{5}{\mu\watt}. The laser power is currently not actively controlled and is thus subject to short term fluctuations and long drifts (aging of the diode). Further testing with the laser power modulated in a controlled fashion will reveal the true correlation between rubidium line frequency and laser power drift. Active control of the laser power or a post-processing correction for power induced shifts in our reference spectrum might become necessary. We also need to find the optimum scan rate to balance intrinsic (photon-limited) SNR and acoustic noise currently limiting the single scan precision.

In order to apply the etalon spectrum for absolute wavelength calibration of a RV spectrograph, we need to determine the exact effective cavity length and the order number of each etalon line in the spectrum. An etalon with dielectric mirror coatings shows a wavelength dependent phase shift (a.k.a. group velocity dispersion) from the wavelength-dependent penetration depth of the light on the mirror surfaces. The amplitude is typically 100--300\,nm over the bandpass of the etalon, which translates into several \si{\kmps} deviation of the true comb line positions versus their ``ideal'' position, i.e. equally spaced in frequency. The group velocity dispersion of the etalon coatings is a known, smooth, and slowly varying function with wavelength \cite{wildi10, schwab15}. We can thus solve for the true line positions or compare the etalon spectrum with a ThAr reference spectrum \cite{bauer15}.

Measuring and tracking our etalon at a single frequency might potentially impose a limitation on applying the etalon comb as a long-term primary wavelength calibrator for high-precision RV measurements. We are relying on the assumption that a drift measured for one etalon line is the same for all comb lines across the whole bandpass, modulo $\lambda_n/\lambda_{ref}$. This is a reasonable assumption, since after stabilizing the input illumination with a SM fiber, the most likely reason for wavelength-dependent effects are changes in the dielectric coatings of the FP etalon mirrors, affecting the penetration depth of the incoming light (and, hence, the cavity length $l$) in a time \textsl{and} wavelength dependent way. Aging and contamination of soft coatings is a known effect, although unlikely to occur for an etalon under vacuum. Nonetheless, minute changes in the coating thickness at the sub-nm level could cause \si{\mps} chromatic line shifts. Hence, simultaneous monitoring at a second reference wavelength or a temporary comparison to a LFC over a moderate amount of time is ultimately desirable to verify the absence of chromatic time-depended effects in a FP etalon. 

\acknowledgments       
 
We acknowledge funding for this project from the David and Lucile Packard Foundation through a fellowship to J.L.B.

\bibliography{report}   
\bibliographystyle{spiejour} 
\end{document}

%% file: units.tex
\DeclareSIUnit{\cmps}{\cm\per\second}
\DeclareSIUnit{\mps}{\meter\per\second}
\DeclareSIUnit{\kmps}{\kilo\meter\per\second}
\DeclareSIUnit{\micron}{\micro\meter}
\DeclareSIUnit{\foot}{'}
\DeclareSIUnit{\inch}{"}